\pdfoutput=1
\documentclass[conference]{IEEEtran}

\usepackage[utf8]{inputenc} 
\usepackage[english]{babel}
\usepackage[T1]{fontenc}
\usepackage{csquotes}
\usepackage{makecell}
\usepackage{cancel}
\usepackage{amsfonts}
\usepackage{graphicx}
\usepackage{subcaption}
\usepackage{textcomp}
\usepackage{float}
\usepackage{algorithm}
\usepackage{algpseudocode}
\usepackage{hyperref}

\usepackage{transparent}

\usepackage{enumitem} 
\usepackage{parskip} 
\usepackage{amsmath}
\usepackage{latexsym}
\usepackage{mathtools}
\usepackage{physics}
\usepackage{amssymb}
\usepackage{comment}
\usepackage{arydshln}

\usepackage{usebib}
\usepackage{comment}
\usepackage{xpatch}

\usepackage{booktabs}   
\usepackage{siunitx}    
\usepackage{adjustbox}  

\usepackage{algpseudocode}
\makeatletter
\let\OldStatex\Statex
\renewcommand{\Statex}[1][3]{%
  \setlength\@tempdima{\algorithmicindent}%
  \OldStatex\hskip\dimexpr#1\@tempdima\relax}
\makeatother

\usepackage{qcircuit}
\usepackage{biblatex}
\hypersetup{colorlinks,%
citecolor=black,%
filecolor=black,%
linkcolor=black,%
urlcolor=black
}
\usepackage{tabularx}
\newcolumntype{P}[1]{>{\hspace{0pt}}p{#1}}
\newcolumntype{K}[1]{>{\hspace{0pt}}X{#1}}
\newcolumntype{L}[1]{>{\raggedright\let\newline\\\arraybackslash\hspace{0pt}}m{#1}}

\newcolumntype{R}[1]{>{\raggedleft\let\newline\\\arraybackslash\hspace{0pt}}m{#1}}
\newcolumntype{C}{>{\centering\arraybackslash}X}

\addto\extrasenglish{}
\addto\extrasenglish{}
\addto\extrasenglish{}

\usepackage{titlesec}
\titleformat{\paragraph}[hang]{\normalfont\normalsize}{\theparagraph}{1em}{}
\titlespacing*{\subsubsection}{0pt}{1.5ex plus 1ex minus .2ex}{0.75em}
\titlespacing*{\paragraph}{0pt}{1.5ex plus 0.75ex minus .2ex}{1em}

\usepackage{tikz,booktabs,array}
\usetikzlibrary{positioning}
\makeatletter
\newcommand{\markZwicky}[1][]{\pgfutil@ifnextchar({\mark@Zwicky{#1}}{\mark@Zwicky{#1}()}}
\def\mark@Zwicky#1(#2)#3{%
   \tikz[every Zwicky picture,#1]{%
     \node[every Zwicky node,draw=none,inner sep=+\z@,outer sep=+\z@] {#3};
     \def\tikz@Mark@name{#2}%
     \ifx\tikz@Mark@name\pgfutil@empty\else
       \tikzset{every Zwicky node/.append style={name={#2}}}%
     \fi
     \node[every Zwicky node,overlay] {\phantom{#3}};
   }%
}
\newcommand{\tikzZwicky}[1][]{%
  \def\tikz@Zwicky@args{#1}%
  \let\tikz@Zwicky@list\pgfutil@gobble
  \let\tikz@Zwicky@first\pgfutil@empty
  \pgfutil@ifnextchar(\tikz@Zwicky@collect\tikz@Zwicky@finish
}
\def\tikz@Zwicky@collect(#1){%
  \ifx\tikz@Zwicky@first\pgfutil@empty
    \edef\tikz@Zwicky@first{#1}%
  \else
    \edef\tikz@Zwicky@list{\tikz@Zwicky@list,#1}%
  \fi
  \pgfutil@ifnextchar(\tikz@Zwicky@collect\tikz@Zwicky@finish
}
\def\tikz@Zwicky@finish{%
  \tikz[remember picture,overlay]
    \draw[every Zwicky connector,/expanded=\tikz@Zwicky@args]
      (\tikz@Zwicky@first) [/expanded={@Zwicky@list/.list={\tikz@Zwicky@list}}] [every Zwicky connect finish/.try];
}
\pgfkeys{/expanded/.code={\edef\pgfkeys@temp{{#1}}\expandafter\pgfkeysalso\pgfkeys@temp}}
\makeatother
\tikzset{
  @Zwicky@list/.style={insert path={to[every Zwicky connector how/.try] (#1)}},
  every Zwicky picture/.style={
    baseline,
    remember picture,
  },
  every Zwicky node/.style={
    remember picture,
    anchor=base,
    inner sep=+2pt
  },
  every Zwicky connector/.style={
    ultra thick,
    red!80!black,
    draw opacity=0.2,
    line cap=round,
    line join=round
  }
}

\makeatletter
\def\@IEEEsectpunct{.\ \,}
\def\paragraph{\@startsection{paragraph}{4}{\z@}{1.5ex plus 1.5ex minus 0.5ex}%
{0ex}{\itshape}}
\makeatother

\usepackage{amsthm}

\begin{document}

\pagestyle{plain}
\pagenumbering{arabic}

\begin{center}
    \LARGE \textbf{Simulating Liquidity: Agent-Based Modeling of Illiquid Markets for Fractional Ownership}
\end{center}

\begin{center}
Lars Fluri\footnote{Universit\"at Basel, Wirtschaftswissenschaftliche Fakult\"at, Peter Merian-Weg~6, 4052 Basel, Email: <\href{mailto:lars.fluri@unibas.ch}{lars.fluri@unibas.ch}>}, A. Ege Yilmaz\footnote{Hochschule Luzern, Institut f\"ur Finanzdienstleistungen Zug IFZ, Suurstoffi~1, 6343 Rotkreuz, Email: <\href{mailto:ahmetege.yilmaz@hslu.ch}{ahmetege.yilmaz@hslu.ch}>}, Denis Bieri\footnote{Hochschule Luzern, Institut f\"ur Finanzdienstleistungen Zug IFZ, Suurstoffi~1, 6343 Rotkreuz, Email: <\href{mailto:denis.bieri@hslu.ch}{denis.bieri@hslu.ch}>}, Thomas Ankenbrand\footnote{Hochschule Luzern, Institut f\"ur Finanzdienstleistungen Zug IFZ, Suurstoffi~1, 6343 Rotkreuz, Email: <\href{mailto:thomas.ankenbrand@hslu.ch}{thomas.ankenbrand@hslu.ch}>}, Aurelio Perucca\footnote{MARK Investment Holding AG, Splint Invest, Unter Altstadt~30, 6300 Zug, Email: <\href{mailto:aurelio.perucca@splintinvest.com}{aurelio.perucca@splintinvest.com}>}
\bigskip
\noindent
\\
\bigskip

\today

\end{center}

\begin{abstract}

This research investigates liquidity dynamics in fractional ownership markets, focusing on illiquid alternative investments traded on a FinTech platform. By leveraging empirical data and employing agent-based modeling (ABM), the study simulates trading behaviors in sell offer-driven systems, providing a foundation for generating insights into how different market structures influence liquidity. The ABM-based simulation model provides a data augmentation environment which allows for the exploration of diverse trading architectures and rules, offering an alternative to direct experimentation. This approach bridges academic theory and practical application, supported by collaboration with industry and Swiss federal funding. The paper lays the foundation for planned extensions, including the identification of a liquidity-maximizing trading environment and the design of a market maker, by simulating the current functioning of the investment platform using an ABM specified with empirical data.





\end{abstract}
\section{Introduction}\label{sec:introduction}

The rapid growth of financial technology has fuelled the emergence of new types of markets, including platforms that enable fractional ownership of traditionally illiquid, unbankable assets such as fine wines, artworks, luxury cars and watches. These markets allow investors to purchase fractional shares of assets, which can then often be traded on secondary markets. However, compared to more liquid and bankable assets such as stocks and bonds, fractional ownership markets face particular liquidity challenges. While economic research has long focused on optimizing market structures for liquid assets, far less attention has been paid to the unique dynamics of fractional ownership markets. Liquidity is a critical factor in any financial market because it determines how easily investors can buy or sell their units and access capital when needed. For markets for fractional ownership, liquidity is influenced by several factors, including the number of overall shares of an asset available, valuation and pricing mechanisms, and the specific rules governing the trading environment. For market providers, accurate modeling of liquidity is critical to understanding trading dynamics and finding ways to improve liquidity. Despite the growing interest in fractionalized assets, research in this area is still underdeveloped. This paper addresses the existing research gap by developing a simulation environment that leverages agent-based models (ABMs) to approximate liquidity patterns on an operational trading platform for fractionalized assets. ABMs are a powerful tool for simulating complex financial markets, particularly those involving heterogeneous agents and varying market conditions. By capturing the behavior and interactions of individual agents, ABMs can offer a more nuanced and accurate representation of market dynamics compared to traditional economic models. Furthermore, ABMs can also be used to generate synthetic data that can represent different market states. This can be valuable since classical econometric or machine learning models usually do not allow for this task of data augmentation. In this study, the ABM framework is applied to simulate the liquidity dynamics of an illiquid secondary market where shares in non-bankable assets are traded. The model is grounded in real-world data provided by a FinTech startup that specializes in fractional ownership of non-bankable assets through its digital platform.


\section{Motivation and Research Question}\label{sec:researchquestion}


This research is motivated by multiple considerations. First, the modeling of non-bankable assets in ABMs is currently not a focal point of research and ABMs for markets without a central liquidity provider are rare, despite their potential in modeling complex systems. Second, the interest in non-bankable, non-traditional assets has surged, primarily driven by the rise of FinTech platforms that are redefining traditional financial services and expanding access to retail investors. Within this broader shift, Distributed Ledger Technology (DLT) and the tokenization of assets are emerging as significant components, granting retail investors access to markets that were previously inaccessible. This leads to a variety of challenges for investors and platform providers. One of the primary challenges is ensuring that these assets are traded on markets with adequate liquidity. Liquidity is essential because it allows investors to buy and sell assets quickly, ensuring portfolio flexibility and access to cash when needed. Investors comparing traditional stock market investments with alternative, non-bankable assets are likely to demand a liquidity premium (or face a discount) for the latter, as less liquid markets expose them to higher risks and delayed exit opportunities. This aligns with \cite{keynes:1936general}, which highlights the importance of liquidity preference, meaning that investors tend to prefer assets that can be easily liquidated over illiquid assets. Further, \cite{tirole:2011} has examined how illiquidity in financial markets can impact asset prices, market stability and overall market efficiency. This is supported by the findings in \cite{amihud:2006}. To increase the attractiveness of their secondary market, platforms offering non-bankables must seek to reduce the impact of lower liquidity on investor decision-making. However, an appropriate understanding of liquidity dynamics is of central importance for this. This research paper simulates such dynamics for an operative platform and lays the basis for identifying potential liquidity-enhancing measures and frameworks. 


\section{Literature Review}\label{sec:literature_review_and_theory}

This section summarizes the state of research of ABMs in finance and more specifically, for the modeling of asset markets. More generally, it highlights key works in financial markets theory relevant to this research. One of the most influential and groundbreaking works in finance is \cite{muth:1961} that describes the concept of rational expectations. \cite{muth:1961} introduced the idea that market participants use all available information to predict future economic conditions, and these predictions typically match actual outcomes,  emphasizing that market participants are usually well-informed and markets efficient. In the advent of progress in behavioral finance, this concept has been increasingly scrutinized. In the research of financial markets, \cite{barclay:2003} state that functioning and therefore liquid markets rely on processes, rules and infrastructures within which the trading of assets takes place. These processes and rules can be summarized as a trading system. A trading system offers market participants a structured framework for interacting with others, facilitating the entry, matching and settlement of orders. \cite{wilensky:2015} described two fundamental forms of trading mechanisms. Order-driven trading implies that participants place buy and sell orders with specific prices and quantities. Those are centrally stored, matched and settled on predefined rules, for example, using continuous trading via central order books or other forms of auctions (e.g. call auctions). Quote-driven trading, in contrast, means that designated market makers or dealers provide prices and quantities for the purchase and sale of securities against which market participants can trade. This trading can take place through centralized entities or over-the-counter (OTC). These types of trading mechanisms have been modeled frequently through the use of ABMs. This is because ABMs give insight into complex systems, i.e. markets, that could otherwise not be obtained using econometric or classic statistical approaches, as stated in \cite{wilensky:2015}. \cite{palmer:1994} provide an ABM of the stock market where independent adaptive agents buy and sell shares of stocks. Those agents select or generate investment rules after observing the outcomes of the mental models adopted in each previous time step. The rules under which their agents act consist of three parts: A condition under which the agent actions are triggered; the action of either buying or selling, and the strength of the optimizing action to increase an agent’s wealth. The agents can react to changing environments. Genetic algorithm (GA) provides evolutionary behavior, i.e. mutation, crossover and selection into the designated
actions. \cite{palmer:1994} generate dynamic and non-equilibrium features, such as bubbles, crashes, wide wealth distribution and higher volatility, which are rarely seen in a typical rational expectations model. \cite{lebaron:2000} provides an overview of six distinct papers that provide important groundwork for the utilization of ABMs in finance. Special focus lies on the heterogeneous behaviors of agents to cross the boundaries of what can be analyzed analytically. ABMs for the specific modeling of trading activities for markets of less liquid assets do not exist in the literature. In traditional economic theory, complex financial markets are modeled using market participants as if they were rational and had homogeneous beliefs. According to \cite{sornette:2009}, these theories cannot model systems close to bifurcation and phase transitions, e.g. bubbles or market crashes and are therefore not realistic. \cite{bouchaud:2008} argues that new models and frameworks are needed in order to give a more realistic representation of financial markets. These models should take into account behavioral aspects. ABMs provide a suitable framework for this purpose. They model markets using microstructures in a so-called bottom-up approach, where agents can have heterogeneous and non-rational behavior. The current research shows multiple research gaps. First, markets with one-sided offer books are not widely explored. Second, low liquidity markets are also not well covered, as most ABM-based research focuses on public stock markets. In the current research, markets are either liquid by design or have a central liquidity provider and market maker that can maintain two-sided quotations, stabilize spreads and ensure trading volumes.  Lastly, ABMs have not yet reached research on markets for fractionalized real assets as most of the current research covers traditional equity markets. The present research paper covers these research gaps in a model that simulates the trading of such typically illiquid assets on a platform with a one-sided offer book and without a central market maker.

\section{Market Setup \& Data}\label{sec:data}

The data upon which the ABM presented in this paper is based are provided by a FinTech startup that specializes in fractional ownership of non-bankable assets such as fine wines, artworks, luxury cars and watches. An asset is fractionalized into initial shares, worth €50 each, and sold to investors via the primary market. The entire asset is held over a pre-defined investment horizon, where individual investors can sell their shares on a secondary market to other users. The present paper aims at simulating this market as accurately as possible by leveraging empirical data for agent identification and specification. The key aspects of the platform's trading mechanisms and rules, i.e. the environment in which market participants (agents) operate, are defined as follows:

\begin{itemize}
    \item \textbf{Market Organization:} The secondary market for trading shares is opened at least once per month.\footnote{Note that the platform adjusted its trading schedule from a four-week cycle to a biweekly window and then to a weekly schedule over the course of the observation period.} Sellers are permitted to list their Shares for sale from 09:00 to 21:00 on designated trading days, while buyers can only make purchases between 18:00 and 21:00 on those same days. Given that buyers are not able to directly place buy orders but can only accept existing sell offers, the market is effectively organized as a sell-side driven offer book. Note that any unmatched sell offers are automatically deleted at the end of each trading day.
    \item \textbf{Trading Rules:} Sell offers are required to be priced within 75\% to 110\% of the asset’s monthly updated reference valuation estimated by the platform provider, ensuring that buyers are shielded from offers that substantially deviate from the current valuation. A 2\% exit fee is charged to sellers upon the completion of a transaction on the secondary market. Buyers do not pay any fees on a successful trade on the secondary market.
    \item \textbf{Order Matching and Market Dynamics:} Sell offers are processed on a first-in, first-out basis by price and listing timestamp. Buyers operate on a first-come, first-served basis, with the system allowing for both exact and partial matches of buy and sell orders.
    \item \textbf{Interactive Trading Interface:} Sellers can adjust or withdraw their offers anytime, while buyers view offers in real-time and make purchasing decisions based on current market conditions.
\end{itemize}
\begin{figure*}[t]
    \centering
    \includegraphics[width=\linewidth]{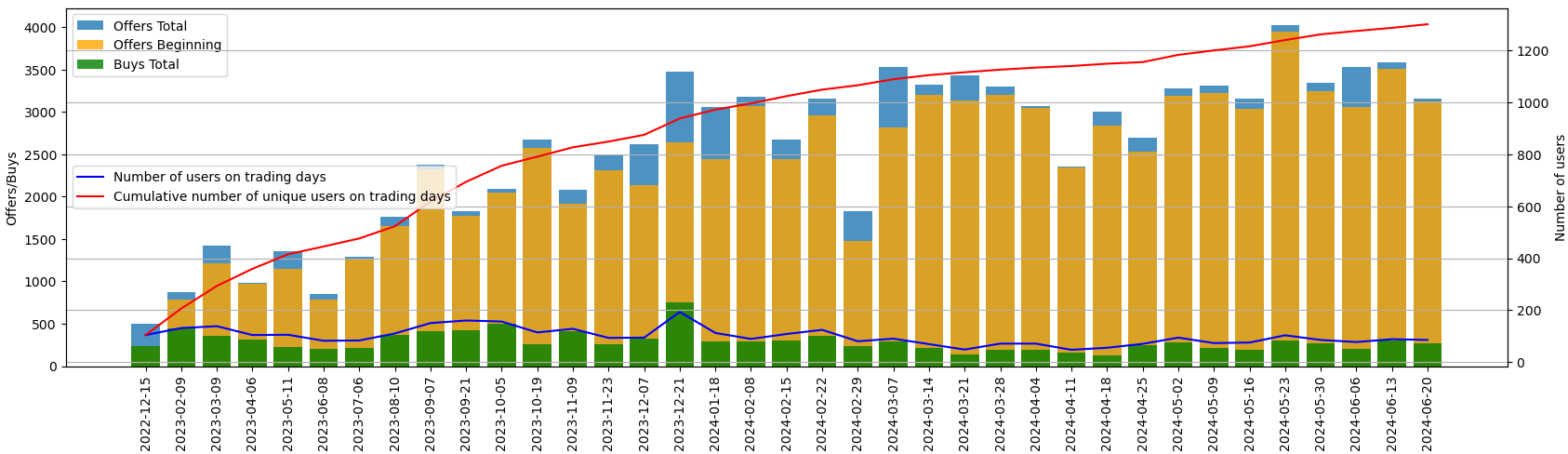}
    \caption{Number of offers and trades shows that most of the offers never get settled. The average liquidity ratio over all trading days is 9.6\%.}
    \label{fig:offers_trades}
\end{figure*}
The dataset provided was subjected to certain pre-processing steps to ensure consistency and alignment with the research question. This included the removal of system accounts, i.e. transactions carried out by the platform itself (e.g. for testing purposes), as well as accounts that were not active on the secondary markets and where a fully consistent replication of all platform activity was not possible (e.g. due to account mergers). The final dataset encompasses 37 regular trading windows spanning from December 2022 to June 2024 (see Figure \ref{fig:offers_trades}). It comprises a total of 26,884 offers and 4,898 matched trades on the secondary market, culminating in a trading volume of €454,330 across 308 distinct assets. Throughout the entire sample period, the platform registered 1,365 individual users that bought and/or sold at least one share of an asset on the secondary market. The empirical data shows that the number of users is relatively stable over the trading windows, while the user base of the platform grew somewhat linearly. The large majority of the sell offers in the offer book is entered during the pre-trading phase. Since the number of offers and trades per trading day are correlated, the liquidity ratio (i.e. the number of traded assets divided by the number of offered assets) remains relatively stable. This stability supports its use as the key performance metric in the simulated market.

Based on the behavior in the secondary market, three distinct agent types for the simulation environment are defined:

\begin{itemize}
    \item \textbf{Pure Buyer (PB)}: This agent type exclusively purchases in the secondary market. The empirical data records 727 such agents.
    \item \textbf{Pure Seller (PS)}: This agent type only sells on the secondary market, having initially bought on the primary market. The empirical data includes 413 of these agents.
    \item \textbf{Buyer Seller (BS)}: This agent type both buys and sells in the secondary market. The empirical data identifies 225 such agents.
\end{itemize}

\autoref{tab:empirical_activity} summarizes the average activity levels derived from the empirical data over all trading windows (TW). It includes the percentage of active participants for the three agent types and trading metrics. While the target metric for evaluating the simulation results is the liquidity ratio, the empirical activity indicators are used to initialize the behaviors of the three agent types. The liquidity ratio is considered a target figure because, from an investor's perspective, the absolute trading volume is not necessarily the most relevant factor. Instead, it is more relevant to assess the proportion of shares offered for sale that actually find a counterparty, as unmatched offers negatively impact platform satisfaction by limiting the user's ability to divest.

\begin{table}[b]
    \centering
     \caption{Average empirical activity levels over all trading days.}
    \begin{tabular}{|l|r|}
        \hline
        \textbf{Metric}                      & \textbf{Average}\\
        \hline
         \textbf{Ratio of active PS} & 0.114 \\
         \textbf{Ratio of active BS (sell-side)}& 0.278 \\
        \textbf{Ratio of active PB}  & 0.092\\
        \textbf{Ratio of active BS (buy-side)}& 0.104 \\
        \textbf{Ratio of Portfolio offered PS} & 0.603\\
        \textbf{Ratio of Portfolio offered BS} &  0.333\\
        \textbf{Number of Trades per TW} & 132\\
        \textbf{Number of Shares traded per TW} & 246\\ 
        \textbf{Platform Revenue per TW (CHF)} &  246\\ 
        \textbf{Liquidity Ratio} & 0.096 \\
        \hline
    \end{tabular}
    \label{tab:empirical_activity}
\end{table}



\section{Methodology}\label{sec:methodology}

This section explains the choice of ABM as the methodological approach for liquidity simulation and provides a general overview of the method. In ABMs, agents are endowed with an action strategy that describes how they interact with other agents. Most of the time, these strategies are linked to an implicit or explicit utility function that describes the underlying preferences of the agents. In the context of finance and financial markets, agents are most often endowed with a budget and a number of assets. They then interact with each other within the given market rules on the basis of their endowments and utility functions. Unlike other methods, which have clearly defined structures and narrow boundaries for creation and parameterization, ABMs provide researchers with a vast array of choice variables. As a result, there is a limited availability of standardized, textbook-like instructional materials for ABMs. \cite{lebaron:2001} offers a general overview and possible design choices for implementing ABMs in a research context.
In general, agents in an ABM can have a variety of setups that can range from quite simple behavior with zero intelligence to rather complex adaptive behavior. For example, \cite{gode:1993} use comparably simple behavior to model agents, while \cite{chen:2001} implement evolving trading behavior using genetic programming as an architecture for agent-based artificial stock markets. Zero-intelligence agents act randomly and are only constrained by their budget.  Another possibility is to make agents able to learn from their environment and then adapt their strategies, as for example in the Santa Fee artificial market (SFI). An implementation of this is showcased in \cite{arthur:2018} and \cite{arifovic:1996}. Another possibility in the context of financial markets is to endow agents with trading rules that seem similar to the real world. This type of modeling of market microstructures is also used in the following simulation approach and was chosen as the predominant and preferred method for two reasons. First, the empirical data provided by the FinTech company actually contains information about the microstructure of the market and the behavior of market participants, which in turn can be used to specify the simulation model.  Using an ABM, a direct comparison between the observed behavior in reality and the simulated behavior is therefore possible. Second, modeling agents individually rather than understanding the market on a global scale allows for adjusting the behavior of the agents on an individual level and then observing new emerging market equilibria. Third, ABMs offer a high degree of flexibility, making it possible to replicate the specific trading conditions of the observed platform. They also facilitate the swift implementation of adjustments to trading conditions and the evaluation of their impact on market equilibrium.

\section{Implementation}\label{sec:implementation}

The concrete specification of the trading conditions on which the simulations are based and the design of the agent behavior are discussed in this section. The components of the model include agent descriptions, the general process of agent interaction and the initialization of the agents. The specific implementation of these components is provided in the form of pseudocode in the \hyperref[sec:algorithms]{Appendix}. The simulation results of the model are described in Section \ref{sec:empirical_results}.
\subsection{Agents}

The description of the agents (decentralized, micro level) includes the inputs for their decision-making, the rules, the variables, the parameters and the output. There are three types of agents: Pure Seller, Pure Buyer and Buyer Seller. These three agents types are defined based on the empirical evidence found in Section \ref{sec:data} and are further detailed in the following paragraphs.

\subsubsection{Pure Seller}

It is assumed that the Pure Seller experiences an exogenous liquidity shock and therefore decides to sell his assets in order to generate liquidity.
With a probability given by the external liquidity shock (PS offer probability), the agent decides to take action, i.e. to offer assets on the market, provided they hold more than zero assets. If the agent opts to act, they offer a quantity based on the offer ratio, defined as the proportion of their total assets they are willing to sell, multiplied by their total assets. The price for each unit of the offered asset is identical and drawn from the uniform distribution
\begin{equation}
    P(p_i) \sim U([v_{PS, l}, v_{PS, h}]\times p_{ref}),
\end{equation}
where $[v_{PS, l}, v_{PS, h}]$ represents the lower and upper bounds of the Pure Seller's price range relative to $p_{ref}$, the reference price, which is the current valuation of the asset, provided by the platform provider in practice. The offer book $O$ is therefore updated with an offer $o_i$, where $o_i$ consists of a price $p_i$, a number of assets $n_i$ and the agent ID of the seller $a_i$. 
\subsubsection{Pure Buyer}

The agent participates in the market given by the Pure Buyer trading probability. The Pure Buyer is not necessarily interested in realizing profits but is instead driven by the desire to acquire assets that they could not purchase in the primary market. The agent is modeled to be a more emotional buyer, that is driven by being partial owner of an asset they desire. Therefore, given that the agent takes action, they look at one randomly picked offer $o_i$ from the offer book $O$ and, depending on the offered price $p_i$, decide whether or not to accept it. The Pure Buyer is more likely to accept lower-priced offers but will still consider and occasionally accept higher-priced ones. The decision to purchase is governed by the probability function
\begin{equation}
    P(p_i) = \frac{1}{1 + e^{k(p_i - p_{ref})}} \quad \forall p_i \in P,
\end{equation}
where $P(p_i)$ is the probability of buying the asset at price $p_i$, $p_{ref}$ is the reference price or current asset valuation and $k$ is the steepness of the decision curve.

The Pure Buyer examines a single offer and then either decides to buy or does not act. If the agent chooses to purchase, they will spend an amount based on their purchase ratio and their available cash. If the cost of the purchased assets is less than their available purchase amount, the Pure Buyer buys all available units $n_{i}$, the offer $o_{i}$ is removed from the offer book and the remaining cash is retained. If the cost of the purchased assets exceeds the available quantity, the agent buys as many units as possible with the available cash, the offer is updated in the order book, and no cash is retained.

\subsubsection{Buyer Seller}

According to the market setup, there is a pre-trading and a trading phase. Since the offer book is predominantly built in the pre-trading phase (see Figure \ref{fig:offers_trades}) and offers get settled in the trading phase, the Buyer Seller agent can be active in both phases. The Buyer Seller aims to maximize profits. During the pre-trading phase, while the offer book is being constructed, the Buyer Seller enters the market with a certain probability (BS offer probability), given their asset balance is positive. If the agent becomes active during this phase, they decide to sell a portion of their assets proportional to the Buyer Seller offer ratio. The price of the assets offered is drawn from the following uniform distribution:
\begin{equation}
    P(p_i) \sim U([v_{BS, l}, v_{BS, h}]\times p_{ref}),
\end{equation}
where $[v_{BS, l}, v_{BS, h}]$ represents the lower and upper bounds of the Buyer Seller's price range relative to $p_{ref}$, the reference price, which is the current valuation of the asset. The offer book $O$ is therefore jointly constructed with the Pure Seller, where each offer $o_i$ consists of a price $p_i$, a corresponding quantity of assets $n_i$ and the seller agent's ID $a_i$. The latter is necessary to ensure that an individual Buyer Seller Agent does not buy their own offers. During the trading phase, the Buyer Seller can also buy assets. Depending on their Buyer Seller search length, the agent looks at a specific amount of randomly picked offers that are priced below the current asset valuation and selects the cheapest among them. This is due to the setup of the platform, where a list-like display of the offer book (or individual assets) is not available. Instead, users can filter for offers below the current asset valuation, with the resulting list not organized by any specific pricing or valuation order. It is therefore reasonable to assume that the agent does not have time to look through all the potential offers. The agent therefore considers a limited amount of offers valued under the current asset valuation and chooses the cheapest one. His decision process is therefore

\begin{equation}
\begin{aligned}
&\min(p_1, p_2, \dots, p_{SL_{BS}}), \\
&\text{where} \quad p_i < p_{ref} \quad \text{for} \quad i = 1, 2, \dots, SL_{BS}.
\end{aligned}
\end{equation}
$SL_{BS}$ is the agent's search length, whereas $p_i$ is the price and $p_{ref}$ is the reference price. The offers are sampled uniformly from all the offers in the offer book below the reference price and do not have to be adjacent. 

The agent then chooses to purchase the offer with the smallest price $p_{min}$. They will spend an amount based on their purchase ratio and their available cash. If the cost of the purchased assets is less than their available purchase amount, the Buyer Seller buys all available units $n_{min}$, the offer $o_{min}$ is removed from the offer book, and the remaining cash is retained. If the cost of the purchased assets exceeds the available quantity, the agent buys as many units as possible with the available cash, the offer is updated in the offer book and no cash is retained.


\subsection{Process}
\label{subsec:process}

The interaction of the agents at the macro level constitutes the collective behaviors and system dynamics that cannot be directly inferred from the actions of individual agents but arise from their local interactions and the aggregate effects of these interactions over time. The corresponding process is described in the following. In line with the current operational rules of the platform, the market is separated into two distinct phases, i.e. pre-trading and trading. The model contains one trading day with pre-trading phase from 09:00 to 18:00 where Pure Sellers and Buyer Sellers can transmit sell offers. The pre-trading phase is modeled in one iteration. This simplification is justified by the circumstance that, according to the empirical data, the vast majority of offers are entered before the actual trading phase (see Figure \ref{fig:offers_trades}). In the second phase, from 18:00 to 21:00, no agents are allowed to place sell offers, and all agents (except the Pure Seller) are permitted to match sell offers from the offer book. The agents in the ABM are endowed with a number of assets and cash derived from the empirical data for the users on the platform. More specifically, the cash position and the aggregated number of assets held are determined for each user on the operating platform, based on activities up to the close of the trading window on 20 June 2024. Thus, the agents' endowments in the ABM represent snapshots from the empirical data. Each simulated agent receives the exact amount of cash and assets as one of its empirical counterparts. In general, agents follow a two-step approach in their actions. In the first step, they decide whether or not to become active in a market. The probability for activation is based on the empirical activity seen in the market. In the second step, assuming they become active, they either sell or buy, based on their underlying decision behavior. The offer and purchase ratios are also derived from the empirical data and defined as average values per agent type. During the trading phase, there are 12 settlement points, each representing a 15-minute interval in practice. During settlement, matched offers get deleted and agent variables get adjusted. The model currently consists of one market.

\subsection{Initialization}
\label{subsec:initialisation}

Parameters that are initialized are the offer probabilities for both Pure Seller and Buyer Seller, offer ratios for both Pure Seller and Buyer Seller, the trading probabilities for the Pure Buyer and the Buyer Seller, the purchase ratios for the Pure Buyer and the Buyer Seller, the decision steepness for the Pure Buyer, the search length for the Buyer Seller, the asset's reference price, the price range of the market and corresponding price ranges for the Pure Seller and Buyer Seller and the number of market iterations. The parameter specification of the baseline model can be found in \autoref{tab:parameter_init}, where the values are defined based on the empirical data for the three agent types.

\begin{table}[t]
    \centering
    \caption{Parameter Specifications for Pure Seller (PS), Pure Buyer (PB) and Buyer Seller (BS) in the baseline model.}
    \begin{tabular}{|l|c|r|}
        \hline
        \textbf{Parameter} & \textbf{Variable} & \textbf{Value in Model}\\
        \hline
        \textbf{PS Offer Probability} &$P_{PS,O}$ & 0.114 \\
        \textbf{PS Offer Ratio} &$R_{PS,O}$& 0.603\\
        \textbf{PS Price Range} &$v_{PS, l}, v_{PS, h}$ &  $[v_{m,l}, (v_{m,h} -0.05)]$ \\
        \textbf{PB Trading Probability} &$P_{PB,T}$& 0.092\\
        \textbf{PB Purchase Ratio} & $R_{PB,P}$& 0.566\\
        \textbf{PB Decision Steepness} & $k_{PB}$ & 2 \\ 
        \textbf{BS Offer Probability} &$P_{BS,O}$& 0.278 \\
        \textbf{BS Offer Ratio} &$R_{BS,O}$& 0.333\\
        \textbf{BS Price Range} &$v_{BS, l}, v_{BS, h}$ & $[v_{m,l}+0.05, (v_{m,h} )]$  \\
        \textbf{BS Trading Probability} & $P_{BS,T}$& 0.104\\
        \textbf{BS Purchase Ratio} & $R_{BS,P}$& 0.485\\
        \textbf{BS Search Length}  & $SL_{BS}$& 5\\
        \textbf{Reference Price} & $P_{ref}$& 50 \\
        \textbf{Price Range Market}&$v_{m,l}, v_{m,h}$ & $[0.75,1.10]$\\
        \textbf{Number of Market Iter.} &$n_{iter,trade}$ & 12 \\
        \hline
    \end{tabular}
    \label{tab:parameter_init}
\end{table}
\section{Empirical Results}
\label{sec:empirical_results}

This section evaluates the simulated liquidity of the baseline model and also conducts a sensitivity analysis to determine how responsive the market model is to changes in the parameters. Corresponding tables can be found in the \hyperref[sec:tables_analysis]{Appendix}.\footnote{Please note that the tables presented mainly relate to the influence of model parameters on overall liquidity measures. Other summary metrics, particularly those at the level of individual agent types, are available from the authors upon request.} The baseline model is evaluated to see whether the simulation can come sufficiently close to the empirical data. The main comparative figure is the liquidity ratio, which, for each market simulation, is computed as 
\begin{equation}
    \text{Liquidity Ratio } = \frac{\text{Total traded Shares}}{\text{Total offered Shares}}.
\end{equation}
If the model cannot approximate the empirical liquidity, further analyses of the parameters and changes to the model architecture would be futile. When comparing Table \ref{tab:baseline_model} to the empirical trading measures provided in Table \ref{tab:empirical_activity}, it can be seen that the model replicates the empirical dynamics sufficiently well. In the baseline model, the average liquidity ratio is 14\%, while the empirical counterpart is at 10\%. Furthermore, the magnitudes of the shares traded in the synthetic versus the empirical market are comparable. 

\begin{table}[b]
\centering
\caption{Metrics for the baseline model}
\begin{tabular}{|l|r|}
\hline
\textbf{Metric} & \textbf{Baseline Model} \\ 
\hline
\textbf{Liquidity Ratio} & 0.139 \\ 
\textbf{Number of Offers} & 69 \\
\textbf{Number of Trades} & 130 \\ 
\textbf{Total Offered Shares} & 4746 \\
\textbf{Total Traded Shares} & 614.28 \\ 
\hline
\end{tabular}
\label{tab:baseline_model}
\end{table}

A sensitivity analysis of the model parameters can give further insights into how variations in agent and market configurations affect market dynamics. For the sensitivity analysis, the ABM was run for 1,000 repeated experiments for the corresponding parameter configurations. In a second step, the average liquidity ratio and numbers of offers as well as trades were calculated.

For the sensitivity analysis, a couple of findings are worth mentioning. Table \ref{tab:sensitivity_analysis_PS} presents the corresponding results of key parameters related to the Pure Sellers (PS) agent type. It shows that the liquidity ratio consistently decreases as the probability of Pure Sellers offering Splints (Shares) ($P_{PS,O}$) increases, as agents put more offers on the market. However, there is also a positive effect on the number of trades and a negative effect on the total number of Splints traded. This is due to a substitution effect. Although more offers mean an increasing number of favorable offers, which increases the total number of trades, these are on average smaller in offer size than the offer sizes of the unchanged Buyer Seller agents. Small trades therefore displace large trades, which leads to a decrease in the total number of Splints sold with an increase in the offer probability of the Pure Seller. An increase in the offer ratio of the Pure Seller ($R_{PS,O}$) also has a negative impact on the liquidity ratio, as the total number of Splints offered grows. An outlier in this respect is the number of offers at a parameter value of 0.4, which is lower than that for higher parameter values. The reason for this is that all Pure Sellers with an asset balance of only few Splints no longer participate in the market, as their offer size is reduced to zero. The inactivity of small Pure Sellers leads to them being substituted by larger sell agents, which has a positive effect on the total number of Splints traded in this specific case. Changes in the lower ($v_{PS,l}$) and upper price bounds ($v_{PS,h}$) have, in tendency, an opposite effect on the liquidity ratio. An increase in the upper bound allows for higher priced offers and consequently leads to fewer trades, negatively impacting the liquidity ratio. While the number of trades also decreases with an increasing lower bound, the total number of settled Splints, however, tends to increase. This in turn is again due to a substitution effect. As Pure Sellers have lower offer sizes on average, these are replaced by larger trades by Buyer Sellers as sellers due to the rising prices. This trend is particularly evident when the lower price bounds of Pure Sellers exceed $0.8$, making their offers more expensive compared to those from Buyer Sellers.

The results for the sensitivity analysis for the Pure Buyer's (PB) parameters are presented in Table \ref{tab:sensitivity_analysis_PB}. It reveals that the liquidity ratio increases as the probability of Pure Buyers trading ($P_{PB,T}$) rises, due to more offers being settled. This is underlined by the increase in the total amount of trades and the total number of Splints settled, while the offer side is unaffected. An increase in the Pure Buyer's purchase ratio ($R_{PB,P}$) has a similar effect. The number of trades increases because the corresponding agents have more money available for trading and therefore have to reject fewer offers due to insufficient funds. Consequently, the higher purchasing power also increases the number of total Splints traded. With regard to the Pure Buyer's decision steepness parameter ($k_{PB}$), no clear effect on liquidity measures is observable.

The results for the sensitivity analysis of the Buyer Seller's (BS) parametrization is shown in Table \ref{tab:sensitivity_analysis_BS}. From the offer perspective, an increase in the Buyer Seller's offer probability ($P_{BS,O}$) and offer ratio ($R_{BS,O}$) impacts the liquidity ratio negatively. This can be explained by an increase in the total number of offers and the total amount of Splints offered in both cases. The rise in the number of offers is directly driven by a higher offer probability, as greater agent activity results in more offers entering the market. Conversely, the decline in the liquidity ratio as the offer ratio increases is attributed to a reduction in the number of inactive Buyer Sellers, who become active also when holding only a small asset balance. It can also be noted that, in addition to the increase in the number of Splints offered, there is also an increase, albeit less pronounced, in the total number of Splints traded when the Buyer Seller offer probability or offer ratio increases. This can be explained by the increased number of offers from Buyer Sellers, which leads to a higher proportion of actual deals from Buyer Sellers on the sell-side compared to the static offer volume from Pure Sellers. Consequently, the budget of Buyer Sellers increases, which strengthens their purchasing power in subsequent iterations of the model and leads to a larger volume of Splints sold. Changes in the Buyer Seller's lower ($v_{BS,l}$) and upper offer price limits ($v_{BS,h}$) tend to have an impact on the liquidity ratio in the same direction, hence, differing from the dynamics for the Pure Seller in an agent comparison. For the Buyer Seller, an increase in both the lower and upper offer price bound decreases the liquidity ratio. This occurs because on average higher offer prices of the Buyer Sellers make their offers less attractive, leading to decreased number of trades and traded Splints. Particularly, an increase in both of the bounds reduces the probability of Pure Buyers accepting offers from Buyer Sellers, leading to fewer trades and a lower amount of total Splints traded. This effect is reinforced by the substitution of larger sell-side Buyer Seller trades with smaller Pure Seller trades, as higher valuation bounds for the Buyer Sellers offers reduces the attractiveness in comparison to those of Pure Sellers. From the buy-side perspective, an increased Buyer Seller trading probability ($P_{BS,T}$) leads to a higher liquidity ratio, as more trades and consequently also a larger number of Splints are settled with unchanged offering dynamics. A similar finding applies to the purchase ratio of the Buyer Seller ($R_{BS,P}$). A corresponding increase also leads to a higher liquidity ratio. The increase in the Buyer Seller's purchasing power works through two channels. On the one hand, trade sizes are increasing directly and on the other hand, there are more trades as insufficient funds are less of a problem for those Buyer Seller's who have only little cash available. A longer search length of the Buyer Seller ($SL_{BS}$) has little influence on the liquidity ratio. This can be explained by the Buyer Seller's buying rule. Since they only see offers below the asset's valuation, they will always accept one, regardless of the number of offers they receive. However, a small increase in the total number of Splints traded can be observed with an increased search length. This can be explained by the fact that an increased number of available offers for Buyer Sellers typically results in lower-priced offers being matched. This leads to a higher number of Splints being traded for a given level of purchasing power or a constant investment budget.

From the offer perspective, further sensitivity analyses are provided in Table \ref{tab:sens_analysis_VR} and Table \ref{tab:sens_analysis_MP}. Table \ref{tab:sens_analysis_VR} evaluates the effect of changing the length of the baseline offering price interval ($\Delta(v_{m,l}, v_{m,h})$) of $[0.75,1.1]$, which translates into alternative ranges within which sellers (i.e. Pure Sellers and Buyer Sellers) can list their assets. It reveals that a larger interval leads, on average, to a similar liquidity ratio. A narrower offering range, in contrast, increases it. This is associated with two effects. First, in the most extreme case illustrated in Table \ref{tab:sens_analysis_VR}, i.e. for a range of $0.15$, only offers below the asset’s valuation are available, meaning that offers from Buyer Sellers are increasingly replaced by those from Pure Sellers in transactions where Buyer Sellers act as buyers. This substitution occurs because Buyer Sellers, following their buying rule, consider only offers below the asset valuation. In the most extreme case, all Pure Seller offers meet this criterion, resulting in a shift toward Pure Sellers on the sell side. Second, the lower maximum price of the offers increases the Pure Buyer's willingness to buy more than the higher minimum price of the offers reduces it, resulting in more trades from the Pure Buyer. The midpoint of the offer price range ($\Tilde{v}_{m}$) is also a relevant factor for the liquidity ratio, as shown in Table \ref{tab:sens_analysis_MP}. The lower the midpoint, and given that the total interval range remains constant, the higher is the liquidity ratio. First, this occurs because the number of sub-valuation offers of the Buyer Seller rises, making Buyer Sellers’ offers increasingly available as potential target trades for other Buyer Sellers. This in turn leads to Buyer Sellers accepting comparatively more offers from the same type of agent, which are also larger than those of the Pure Sellers on average. Second, a lower midpoint implies a higher trading probability of Pure Buyers, resulting in more trades and Splints settled.

\section{Conclusion}
\label{sec:conclusion}

This research shows how illiquid markets with non-bankable assets can be modeled using an agent-based model and how different market parameterizations influence the market liquidity.  The results show that the simulated experiments, as measured by the market liquidity and the magnitude of trades, are close to their empirical counterparts. Therefore, the simulation environment can be used to simulate various market regimes and rules. This is valuable to determine liquidity drives and liquidity-maximizing market regimes.
It can be seen from the results that the ABM works as required in the following ways. First, the model is relatively stable. This means that while the liquidity does change for different parameterizations, the model, on average, does not completely change in terms of liquidity and other metrics. Second, changing parameters indeed do have an influence on market liquidity. In particular, the sensitivity analysis reveals several expected and unexpected effects of parameter changes on market liquidity. While some outcomes, such as increased liquidity ratios with higher buying activity and purchasing power, align with intuition, other effects are more counterintuitive. For instance, the substitution of larger trades for smaller ones as the Pure Seller offer probability rises leads to an increase in the total number of Splints traded, even though the number of trades decreases. Additionally, the interplay between price bounds and liquidity dynamics demonstrates that adjustments to lower and upper offer price limits can have opposing effects, depending on the agent type and market context. These findings emphasize that liquidity optimization requires careful consideration of not only direct effects but also more complex interactions within the market structure. In summary, the model implementation can be used for the study of liquidity within illiquid markets and furthermore also provides the possibility of data augmentation based on synthetic data from an agent-based model. It therefore presents a novel model environment for illiquid assets in fractional ownership markets and can be used to generate additional insights into market liquidity optimization. It is characterized by multiple unique traits: It is based on a market with a one-sided offer book, models fractionalized assets and does not implement a central market maker or liquidity provider. This uniqueness makes it a valuable addition to current research and also allows for valuable extensions, as will be described in Subsection \ref{subsec:extensions}.

\subsection{Limitations}
\label{subsec:limitations}
This paper is limited by various factors. First, the implementation is market-specific and does not necessarily extend to other market environments directly.
Since some parameterization relies on empirical counterparts or equivalents, this means that different markets will elicit different parameterizations.
Furthermore, the market and trading mechanisms are also modeled to closely resemble the empirical counterparts and therefore do not generalize. 
This limitation is not specific to this research paper and also applies to various other projects using ABM: While these models are great for modeling market microstructures, they are usually specific to the use case and therefore cannot be directly generalized to other markets.
\subsection{Extensions}
\label{subsec:extensions}

The roadmap for extending this research includes two main steps: First, enhancing agent behavior diversity by incorporating heterogeneous behaviors alongside varied endowments (shares and budgets) and applying regression analysis on simulated liquidity data to quantitatively identify the factors driving liquidity. Second, exploring changes in the market environment, including the impact of different market regimes (e.g. two-sided order books) and the introduction of a market maker agent, to identify configurations that maximize liquidity. This approach enables comparisons with existing ABM literature, refines understanding of how different market structures influence liquidity, and offers insights for the platform provider to adjust its platform design to enhance liquidity in the secondary market.


\printbibliography
\onecolumn

\section*{Appendix: Baseline Results and Sensitivity Analysis}\label{sec:tables_analysis}

\begin{table*}[h]
    \centering
    \caption{Sensitivity analysis of PS parameters over 1,000 repeated experiments; results describe averages.}
    \label{tab:sensitivity_analysis_PS}
    \begin{tabular}{|l|r|r|r|r|r|}
        \hline
        \textbf{$P_{PS,O}$} & 0.01 & 0.06 & 0.11 & 0.16 & 0.21 \\
        \hline
        \textbf{Liquidity Ratio} & 0.179 & 0.157 & 0.14 & 0.123 & 0.113 \\
        \textbf{Number of Offers} & 47.677 & 58.145 & 68.832 & 79.292 & 89.904 \\
        \textbf{Number of Trades} & 124.751 & 127.912 & 130.332 & 132.366 & 134.239 \\
        \textbf{Total Offered Shares} & 4042.423 & 4395.89 & 4714.857 & 5163.784 & 5508.248 \\
        \textbf{Total Traded Shares} & 670.363 & 643.477 & 618.244 & 601.727 & 593.602 \\
        \hline
        $R_{PS,O}$ & 0.4 & 0.5 & 0.6 & 0.7 & 0.8 \\
        \hline
        \textbf{Liquidity Ratio} & 0.153 & 0.142 & 0.141 & 0.135 & 0.131 \\
        \textbf{Number of Offers} & 63.504 & 68.742 & 68.783 & 68.652 & 68.629 \\
        \textbf{Number of Trades} & 129.888 & 130.391 & 130.759 & 130.36 & 130.234 \\
        \textbf{Total Offered Shares} & 4432.697 & 4659.107 & 4715.071 & 4893.038 & 5019.489 \\
        \textbf{Total Traded Shares} & 635.894 & 613.337 & 618.741 & 618.715 & 621.908 \\
        \hline
        $v_{PS,l}$ & 0.55 & 0.65 & 0.75 & 0.85 & 0.95 \\
       \hline
        \textbf{Liquidity Ratio} & 0.137 & 0.136 & 0.137 & 0.145 & 0.144 \\
\textbf{Number of Offers} & 68.712 & 68.832 & 69.115 & 68.436 & 68.833 \\
\textbf{Number of Trades} & 135.925 & 133.357 & 130.918 & 127.451 & 120.654 \\
\textbf{Total Offered Shares} & 4753.666 & 4799.024 & 4761.894 & 4755.109 & 4740.565 \\
\textbf{Total Traded Shares} & 611.911 & 611.38 & 613.126 & 642.282 & 640.64 \\
\hline
        $v_{PS,h}$ & 0.9 & 1.0 & 1.1 & 1.2 & 1.3 \\
        \hline
        \textbf{Liquidity Ratio} & 0.14 & 0.139 & 0.136 & 0.134 & 0.13 \\
        \textbf{Number of Offers} & 68.411 & 68.712 & 68.403 & 69.05 & 68.496 \\
        \textbf{Number of Trades} & 136.27 & 134.688 & 126.606 & 120.716 & 116.674 \\
        \textbf{Total Offered Shares} & 4709.046 & 4753.892 & 4735.58 & 4709.53 & 4720.215 \\
        \textbf{Total Traded Shares} & 614.743 & 625.265 & 604.201 & 592.235 & 579.391 \\
        \hline
    \end{tabular}
\end{table*}

\begin{table*}[h]
    \centering
    \caption{Sensitivity analysis of PB agent over 1,000 repeated experiments; results describe averages.}
    \label{tab:sensitivity_analysis_PB}
    \begin{tabular}{|l|r|r|r|r|r|}
        \hline
        $P_{PB,T}$ & 0.01 & 0.05 & 0.09 & 0.13 & 0.17 \\
        \hline
        \textbf{Liquidity Ratio} & 0.071 & 0.107 & 0.139 & 0.167 & 0.19 \\
        \textbf{Number of Offers} & 68.602 & 68.647 & 68.755 & 68.451 & 69.022 \\
        \textbf{Number of Trades} & 69.35 & 101.174 & 131.4 & 156.174 & 180.286 \\
        \textbf{Total Offered Shares} & 4689.23 & 4743.546 & 4692.506 & 4686.342 & 4757.278 \\
        \textbf{Total Traded Shares} & 312.309 & 477.011 & 611.051 & 731.432 & 847.914 \\
        \hline
        $R_{PB,P}$ & 0.37 & 0.47 & 0.57 & 0.67 & 0.77 \\
        \hline
        \textbf{Liquidity Ratio} & 0.114 & 0.129 & 0.14 & 0.152 & 0.159 \\
        \textbf{Number of Offers} & 68.339 & 68.398 & 68.517 & 68.66 & 68.534 \\
        \textbf{Number of Trades} & 114.106 & 125.223 & 130.872 & 135.238 & 139.738 \\
        \textbf{Total Offered Shares} & 4697.465 & 4703.319 & 4675.799 & 4656.446 & 4752.768 \\
        \textbf{Total Traded Shares} & 503.411 & 571.602 & 614.618 & 660.319 & 705.514 \\
        \hline
        $k_{PB}$ & 0.5 & 1.5 & 2.0 & 2.5 & 3.0 \\
        \hline
         \textbf{Liquidity Ratio} & 0.136 & 0.139 & 0.138 & 0.136 & 0.137 \\
        \textbf{Number of Offers} & 68.527 & 68.74 & 69.033 & 68.715 & 68.75 \\
        \textbf{Number of Trades} & 128.377 & 129.845 & 130.496 & 130.547 & 130.219 \\
        \textbf{Total Offered Shares} & 4718.214 & 4720.788 & 4789.39 & 4782.177 & 4759.494 \\
        \textbf{Total Traded Shares} & 599.643 & 615.622 & 620.542 & 615.977 & 610.49 \\
        \hline
    \end{tabular}
\end{table*}

\begin{table*}[h]
    \centering
    \caption{Sensitivity analysis of BS agent over 1,000 repeated experiments; results describe averages.}
    \label{tab:sensitivity_analysis_BS}
    \begin{tabular}{|l|r|r|r|r|r|}
        \hline
        $P_{BS,O}$ & 0.08 & 0.18 & 0.28 & 0.38 & 0.48 \\
        \hline
         \textbf{Liquidity Ratio} & 0.33 & 0.196 & 0.137 & 0.107 & 0.09 \\
        \textbf{Number of Offers} & 36.11 & 52.207 & 68.443 & 84.755 & 101.417 \\
        \textbf{Number of Trades} & 128.4 & 129.799 & 129.977 & 130.528 & 130.147 \\
        \textbf{Total Offered Shares} & 1927.771 & 3306.697 & 4730.102 & 6181.488 & 7501.16 \\
        \textbf{Total Traded Shares} & 545.08 & 589.1 & 606.98 & 635.524 & 653.342 \\
        \hline 
        $R_{BS,O}$ & 0.13 & 0.23 & 0.33 & 0.43 & 0.53 \\
        \hline
        \textbf{Liquidity Ratio} & 0.253 & 0.179 & 0.139 & 0.114 & 0.097 \\
        \textbf{Number of Offers} & 64.703 & 66.968 & 68.959 & 70.224 & 73.176 \\
        \textbf{Number of Trades} & 132.878 & 131.482 & 130.507 & 130.225 & 130.174 \\
        \textbf{Total Offered Shares} & 2332.204 & 3517.687 & 4690.474 & 5927.671 & 7108.018 \\
        \textbf{Total Traded Shares} & 553.473 & 590.684 & 616.22 & 631.673 & 641.739 \\
        \hline
        $P_{BS,T}$ & 0.01 & 0.05 & 0.1 & 0.15 & 0.2 \\
        \hline
         \textbf{Liquidity Ratio} & 0.079 & 0.106 & 0.138 & 0.169 & 0.197 \\
        \textbf{Number of Offers} & 68.968 & 68.867 & 68.396 & 68.543 & 68.629 \\
        \textbf{Number of Trades} & 77.061 & 102.138 & 130.589 & 155.614 & 178.937 \\
        \textbf{Total Offered Shares} & 4715.668 & 4779.289 & 4779.716 & 4706.767 & 4735.425 \\
        \textbf{Total Traded Shares} & 345.774 & 475.418 & 617.628 & 747.772 & 875.134 \\
        \hline
         $R_{BS,P}$& 0.29 & 0.39 & 0.49 & 0.59 & 0.69 \\
        \hline 
        \textbf{Liquidity Ratio} & 0.114 & 0.126 & 0.139 & 0.149 & 0.161 \\
        \textbf{Number of Offers} & 68.206 & 68.827 & 68.616 & 68.466 & 68.537 \\
        \textbf{Number of Trades} & 118.921 & 123.608 & 130.278 & 133.821 & 136.4 \\
        \textbf{Total Offered Shares} & 4750.504 & 4740.672 & 4766.809 & 4749.974 & 4692.752 \\
        \textbf{Total Traded Shares} & 504.64 & 557.728 & 621.006 & 665.05 & 707.104 \\
        \hline
        $SL_{BS}$ & 1 & 3 & 5 & 7 & 9 \\
        \hline
        \textbf{Liquidity Ratio} & 0.14 & 0.14 & 0.14 & 0.14 & 0.14 \\
        \textbf{Number of Offers} & 68.61 & 68.57 & 68.66 & 68.80 & 68.49 \\
        \textbf{Number of Trades} & 130.05 & 130.03 & 130.64 & 130.94 & 130.35 \\
        \textbf{Total Offered Shares} & 4664.96 & 4733.36 & 4774.73 & 4717.79 & 4800.83 \\
        \textbf{Total Traded Shares} & 600.94 & 606.69 & 613.88 & 624.93 & 634.76 \\
        \hline
        $v_{BS,l}$ & 0.55 & 0.65 & 0.75 & 0.85 & 0.95 \\
        \hline
        \textbf{Liquidity Ratio} & 0.187 & 0.168 & 0.149 & 0.129 & 0.096 \\
        \textbf{Number of Offers} & 68.283 & 68.56 & 68.889 & 68.28 & 68.456 \\
        \textbf{Number of Trades} & 153.61 & 145.332 & 135.733 & 124.516 & 105.013 \\
        \textbf{Total Offered Shares} & 4663.728 & 4672.436 & 4710.982 & 4690.606 & 4710.157 \\
        \textbf{Total Traded Shares} & 816.745 & 740.779 & 657.338 & 567.286 & 425.707 \\
        \hline 
        $v_{BS,h}$ & 0.9 & 1.0 & 1.1 & 1.2 & 1.3 \\
        \hline
        \textbf{Liquidity Ratio} & 0.185 & 0.169 & 0.14 & 0.123 & 0.111 \\
        \textbf{Number of Offers} & 68.556 & 68.659 & 68.441 & 68.151 & 68.408 \\
        \textbf{Number of Trades} & 154.171 & 150.297 & 130.286 & 117.841 & 109.727 \\
        \textbf{Total Offered Shares} & 4740.458 & 4695.305 & 4703.409 & 4790.82 & 4745.347 \\
        \textbf{Total Traded Shares} & 815.064 & 740.42 & 620.561 & 547.259 & 494.064 \\
        \hline
    \end{tabular}
\end{table*}

\begin{table*}[h]
\centering
\caption{Sensitivity analysis of market parameters with varying valuation range over 1,000 repeated experiments; results describe averages.}
\label{tab:sens_analysis_VR}
\begin{tabular}{|l|r|r|r|r|}
\hline
$v_{m,l}$, $v_{m,h}$ & 0.550, 1.300 & 0.650, 1.200 & 0.750, 1.100 & 0.850, 1.000 \\
$\Delta(v_{m,l}, v_{m,h})$ & 0.750 & 0.550 & 0.350 & 0.150 \\
\hline
\textbf{Liquidity Ratio} & 0.142 & 0.139 & 0.139 & 0.157 \\
\textbf{Number of offers} & 68.312 & 68.267 & 68.633 & 68.790 \\
\textbf{Number of Trades} & 126.010 & 125.941 & 130.356 & 151.282 \\
\textbf{Total offered Shares} & 4747.753 & 4691.659 & 4783.819 & 4740.028 \\
\textbf{Total Traded Shares} & 635.311 & 612.775 & 619.830 & 697.677 \\
\hline
\end{tabular}
\end{table*}

\begin{table*}[h]
\centering
\caption{Metrics by valuation range with changing midpoints over 1,000 repeated experiments; results describe averages.}
\label{tab:sens_analysis_MP}
\begin{tabular}{|l|r|r|r|r|r|}
\hline
$v_{m,l}$ & 0.675 & 0.750 & 0.825 & 0.900 & 0.975 \\
$v_{m,h}$ & 1.025 & 1.100 & 1.175 & 1.250 & 1.325 \\
$\Tilde{v}_{m}$ & 0.850 & 0.925 & 1.000 & 1.075 & 1.150 \\
\hline
\textbf{Liquidity Ratio} & 0.174 & 0.139 & 0.107 & 0.074 & 0.012 \\
\textbf{Number of Offers} & 68.677 & 68.939 & 68.418 & 68.238 & 68.538 \\
\textbf{Number of Trades} & 157.925 & 130.409 & 101.637 & 71.484 & 12.515 \\
\textbf{Total Offered Shares} & 4773.591 & 4706.489 & 4685.683 & 4796.811 & 4785.100 \\
\textbf{Total Traded Shares} & 773.682 & 614.081 & 470.680 & 332.259 & 55.455 \\
\hline
\end{tabular}
\end{table*}

\clearpage

\section*{Appendix: Algorithms}\label{sec:algorithms}


\begin{algorithm}
    \caption{Pure Seller's (PS) decision to act and offer shares for one PS agent $a_i$}
    \begin{algorithmic}[1]
        \Require None
        \Ensure An offer is entered into the offer book $O$, or no action is taken
        \State \textbf{Variables:} Shares $s_i$
        \State \textbf{Parameters:}
                $P_{\text{PS},O}$, $R_{\text{PS},O}$, $v_{\text{PS},l}$, $v_{\text{PS},h}$, $p_{\text{ref}}$
        \If{$U(0,1) < P_{\text{PS},O}$ \textbf{and} $s_i > 0$}
            \State $n_i \leftarrow \lfloor R_{\text{PS},O} \times s_i \rfloor$ \Comment{Compute number of shares to offer}
            \State $p_i \leftarrow U(v_{\text{PS},l} \times p_{\text{ref}},\; v_{\text{PS},h} \times p_{\text{ref}})$ \Comment{Set offer price uniformly within price range}
            \State $o_i = (p_i, n_i, a_i)$ \Comment{Create offer}
            \State $O \leftarrow O \cup \{o_i\}$ \Comment{Add offer to offer book}
        \Else
            \State No action taken
        \EndIf
    \end{algorithmic}
\end{algorithm}

\begin{algorithm}
    \caption{Pure Buyer's (PB) decision to act and buy shares for one PB agent $a_j$ and one seller agent $a_i$}
    \begin{algorithmic}[1]
        \Require Offer book $O$
        \Ensure Accept an offer from the offer book $O$ or do nothing
        \State \textbf{Variables:} Shares $s_j$, cash $c_j$, Shares $s_i$, cash $c_i$
        \State \textbf{Parameters:}
                $P_{\text{PB},T}$, $R_{\text{PB},P}$, $p_{\text{ref}}$, $k_{\text{PB}}$

        \If{$U(0,1) < P_{\text{PB},T}$}
            \State Randomly select an offer $o_i = (p_i, n_i, a_i)$ from $O$
            \State $P(p_i) \leftarrow \frac{1}{1 + e^{k_{\text{PB}}( p_i - p_{\text{ref}})}}$ \Comment{Calculate purchase probability}
            \If{$U(0,1) < P(p_i)$} \Comment{PB acts as buyer}
                \State $V_p \leftarrow R_{\text{PB},P} \times c_j$ \Comment{Compute purchase volume}
                \If{$V_p \geq p_i \times n_i$} \Comment{PB can purchase full offer}
                    \State $s_j = s_j + n_i$ \Comment{Update PB's shares}
                    \State $c_j = c_j - (p_i \times n_i)$ \Comment{Update PB's cash}
                    \State $s_i =s_i - n_i$ \Comment{Update seller's shares}
                    \State $c_i = c_i + (p_i \times n_i)$ \Comment{Update seller's cash}
                    \State $O \leftarrow O \setminus \{o_i\}$ \Comment{Remove offer from offer book}
                \Else \Comment{PB can only purchase a fraction of the offer}
                    \State $n_p \leftarrow \left\lfloor \frac{V_p}{p_i} \right\rfloor$  \Comment{Compute purchase quantity}
                    \If{$n_p > 0$} \Comment{Quantity is non-zero}
                           \State $s_j = s_j + n_p$ \Comment{Update PB's shares}
                           \State $c_j = c_j - (p_i \times n_p)$ \Comment{Update PB's cash}
                           \State $s_i = s_i - n_p$ \Comment{Update seller's shares}
                           \State $c_i = c_i + (p_i \times n_p)$ \Comment{Update seller's cash}
                           \State $n_i \leftarrow n_i - n_p$ \Comment{Update remaining offer quantity}
                           \State $O \leftarrow O$ with $o_{i}$ updated \Comment{Update offer in offer book}
                    \Else
                        \State No action taken (insufficient funds to purchase any shares)
                    \EndIf
                \EndIf
            \Else
                \State No action taken (offer not accepted)
            \EndIf
        \Else
            \State No action taken during trading phase
        \EndIf
    \end{algorithmic}
\end{algorithm}

\begin{algorithm}
    \caption{Buyer Seller's (BS) decision to act and offer and/or buy shares for one BS agent $a_{i,BS}$ and one seller agent $a_{min}$}
    \begin{algorithmic}[1]
        \Require Offer book $O$
        \Ensure Enter an offer in the offer book, accept an offer from the offer book, or do nothing
        \State \textbf{Variables:} Shares $s_{i,BS}$, cash $c_{i,BS}$, Shares $s_{min}$, cash $c_{min}$
   
        \State \textbf{Parameters:} $P_{BS,O}$, $R_{BS,O}$, $v_{BS,l}$, $v_{BS,h}$, $p_{ref}$, $P_{BS,T}$, $SL_{BS}$, $R_{BS,P}$
        \State \textbf{Pre-trading phase} (BS acts as a seller):
        \If{$U(0,1) < P_{BS,O}$ \textbf{and} $s_{i,BS} > 0$}
            \State $n_{i} \leftarrow\lfloor R_{BS,O} \times s_{i,BS}\rfloor$ \Comment{Compute number of shares to offer}
            \State $p_{i} \leftarrow U(v_{BS,l} \times p_{ref}, v_{BS,h} \times p_{ref})$ \Comment{Set offer price uniformly within price range}
            \State $o_{i} = (p_{i}, n_{i}, a_{i,BS})$ \Comment{Create offer}

            \State $O \leftarrow O \cup \{o_{i}\}$ \Comment{Add offer to offer book}
        \Else
            \State No action taken during pre-trading phase
        \EndIf
        
        \State \textbf{Trading phase} (BS acts as a buyer):
        \If{$U(0,1) < P_{BS,T}$}
            \State Filter offers $O_f$ from $O$ where $p_i < p_{ref}$ \textbf{and} $a_{i} \neq a_{i,BS}$ \Comment{Select non-own offers with price below reference price}
            \If{$|O_f| > 0$} \Comment{Set of offers is not empty}
                \State Randomly sample $SL_{BS}$ offers from $O_f$ \Comment{Sample offers based on search length}
                \State Choose the offer with the minimum price $o_{min} = (p_{min}, n_{min}, a_{min})$ \Comment{Select the cheapest offer}
                \State $V_p \leftarrow R_{BS,P} \times c_{i,BS}$ \Comment{Compute purchase volume}
                \If{$V_p \geq p_{min} \times n_{min}$} \Comment{PS can purchase full offer}
                        \State $s_{i,BS} \leftarrow s_{i,BS} + n_{min}$  \Comment{Update BS's shares}
                        \State $c_{i,BS} \leftarrow c_{i,BS} - (p_{min} \times n_{min})$  \Comment{Update BS's cash}
                        \State $s_{min} \leftarrow s_{min} - n_{min}$  \Comment{Update seller's shares}
                        \State $c_{min} \leftarrow c_{min} + (p_{min} \times n_{min})$  \Comment{Update seller's cash}
                    \State $O \leftarrow O \setminus \{o_{min}\}$ \Comment{Remove offer from offer book}
                \Else \Comment{BS can only purchase a fraction of the offer}
                    \State $n_p \leftarrow \left\lfloor \frac{V_p}{p_{min}} \right\rfloor$  \Comment{Compute purchase quantity}
                    \If{$n_p > 0$} \Comment{Quantity is non-zero}
                            \State $s_{i,BS} \leftarrow s_{i,BS} + n_p$  \Comment{Update BS's shares}
                            \State $c_{i,BS} \leftarrow c_{i,BS} - (p_{min} \times n_p)$ \Comment{Update BS's cash}
                            \State $s_{min} \leftarrow s_{min} - n_p$ \Comment{Update seller's shares}
                            \State $c_{min} \leftarrow c_{min} + (p_{min} \times n_p)$ \Comment{Update seller's cash}
                        \State $n_{min} \leftarrow n_{min} - n_p$ \Comment{Update remaining offer quantity}
                        \State $O \leftarrow O$ with $o_{min}$ updated \Comment{Update offer in offer book}
                    \Else
                        \State No action taken (insufficient funds to purchase any shares)
                    \EndIf
                \EndIf
            \Else
                \State No offers with prices below reference price available
            \EndIf
        \Else
            \State No action taken during trading phase
        \EndIf
    \end{algorithmic}
\end{algorithm}

\begin{algorithm}
    \caption{General process of the pre-trading and trading phases}
    \begin{algorithmic}[1]
        \Require Set of agents $A$, parameters $\Theta$
        \Ensure Summary metrics $M$
        \State \textbf{Initialization:} Equip agents with values for variables and parameters
        \For{each agent $a \in A$ \textbf{where} $a$ is a \textbf{Pure Seller} or \textbf{Buyer Seller}}
            \State Determine offer action based on $P_{PS,O}$ or $P_{BS,O}$, depending on the agent type
            \State Generate offer $o(a)$ if the action results in an offer
        \EndFor
        \For{each time step $t = 1$ to $T$}
            \For{each agent $a \in A$ \textbf{where} $a$ is a \textbf{Pure Buyer} or \textbf{Buyer Seller}}
                \State Determine trading action based on $P_{PB,T}$ or $P_{BS,T}$, depending on the agent type
                \State Match offer between buyer and seller if the action results in a purchase
                \State Settle transaction and update agent states
            \EndFor
            \State Update summary metrics $M(t)$
        \EndFor
    \end{algorithmic}
\end{algorithm}

\end{document}